# Investigation of Factors Affecting Vertical Sag of Stretched Wire

Jiandong Yuan[1,2,*], Junxia Wu[1], Bin Zhang[1], Yuan He[1], Junhui Zhang[1], Wenjun Chen[1], Shaoming Wang[1], Guozhen Sun[1], Xundong Zhang[1], Lisong Yan[1]

[1]Institute of Modern Physics, Chinese Academy of Sciences, Lanzhou, 730000, China

[2]School of Nuclear Science and Technology, University of Chinese Academy of Sciences, Beijing, 100049, China

**Abstract:** To study vertical sag requirements and factors affecting the stretched wire alignment method, the vertical sag equation is first derived theoretically. Subsequently, the influencing factors (such as the hanging weight or tension, span length, temperature change, elastic deformation, and the Earth's rotation) of the vertical sag are summarized, and their validity is verified through actual measurements. Finally, the essential factors affecting vertical sag, i.e., the specific strength and length, are discussed. It is believed that the vertical sag of a stretched wire is proportional to the square of the length and inversely proportional to the specific strength of the material.

**Keywords:** stretched wire; alignment; vertical sag; catenary; hyperbolic cosine; linear density; Earth rotation

## 1 Introduction

The sag of a stretched wire is also known as a vertical vector, vertical distance, or deflection and directly determines the observation quality and reliability of a stretched wire [1]. To simultaneously improve the length and accuracy of a stretched wire, the smallest vertical sag is typically desired. For example, in various accelerator alignment measurements [2], such as the calibration of beam diagnosis components [3] and various types of magnets [2], numerous Be–Cu wires of diameter 0.1–0.5 mm and length 2–10 m are used, and the vertical sag is required to be less than 20 μm [1]. Meanwhile, in water conservancy and hydropower monitoring, where the distance involved is typically greater than 100 m, high-strength stainless steel of diameter 0.5–1.5 mm is often used to lay stretched wires, and the vertical sag is typically ≤ 0.5 m. Otherwise, the protective tube will be extremely large (at least greater than 1 m), which is unfavorable for the construction and maintenance of dams [4]. In fact, the vertical sag of a stretched wire is proportional to its length; therefore, the longer its length, the larger is its vertical sag. Although increasing the tension/hanging weight can reduce the vertical sag, the risk of wire breakage is

*Corresponding author. *E-mail address:* yuanjiandong@impcas.ac.cn

Large Research Infrastructures "China initiative Accelerator Driven System" (Grant No.2017-000052-75-01-000590 )

increased [5]. Therefore, it is valuable to determine the reasonable vertical sag of a stretched wire. Moreover, because of the high performance and practicability of the stretched wire alignment method, the vertical sag requirements should account for the accuracy requirements and breaking strength of the material.

Although it has been established that a catenary can accurately describe the vertical sag of a stretched wire, as early as 1638, Galileo used a parabola to analyze a stretched wire in his book, ***"Two New Sciences."*** [6] Until the 1690s, Leibniz, Huygens, etc. considered appropriate models of the hyperbolic cosine catenary [7,8]. In 1697, Bernoulli used calculus to solve the catenary as a hyperbolic cosine [8]. Wang et al. in 2014 [9], Zhang et al. [10] and He et al. [11] in 2016, Stanton et al. [12] in 2003, and A. Temnykh et al. [13] in 2010 deduced the vertical sag equation. Furthermore, many excellent studies have been published regarding the vertical sag measurement of a stretched wire. For example, Sudou et al. [14] used a Charge Coupled Device (CCD) in 1996 to achieve high-precision noncontact measurement of wire sagging in an electrostatic drift chamber; however, each point required a combination of two cameras to achieve two-dimensional measurements. Moreover, the measuring range was only $\pm 1$ mm, and the depth of field was only 1 mm; as such, many CCD cameras were required, and it was impossible to measure the vertical sag of all the wire positions. In 2008, Lan et al. [15] utilized a CCD to measure the vertical sag of a stretched wire. In 2015, Wang et al. [16] used the contact-type Light Emitting Diode (LED) measurement method to measure the vertical sag of any point; however, the connection to the wire caused a significant measurement error. In 2016, Zhang et al. [10] vibrated a wire and measured its frequency to calculate the vertical sag of a wire. However, the quantitative selection of vertical sag for different materials to ensure the nonbreakage of wire [5] remained unsolved.

To solve this problem, we first derived an equation to calculate the maximum vertical sag based on a rederived catenary equation. Subsequently, we analyzed its influencing factors and derived a unified equation for the maximum vertical sag. Next, we designed and established a stretch equipment and used a Keuffel and Esser level to measure the vertical sag of stainless steel 304 and Be–Cu wires of different radii. We demonstrate that the material radius or diameter did not directly affect the vertical sag of the stretched wire but only affects the maximum tensile force that the material can withstand. In the discussion section, the maximum tensile force and specific strength are analyzed. The results show that the vertical sag is proportional to the square of the length and inversely proportional to the specific strength of the material.

## 2 Principle Overview

The stretched wire method is one of many baseline/direction line measurements; it is also known as the alignment, string, or micro-offset measurement [1,2] method. A mechanically suspended wire is generally used at two fixed positions on the ground to establish a nondisturbed measurement reference line. An appropriate weight is applied to maintain the straightness of the wire and then used to measure a series of one-dimensional or two-dimensional microscopic horizontal and/or vertical offset of points to be inspected [1,2]. Even if the fixed end and the tension end of the stretched wire are installed on a stable foundation, and the measurement is performed indoors without wind, the stretched wire will still sag under the combined effect of the hang tension and the wire's dead weight. Therefore, the calculus presented in previous studies [17-20] was used to derive an equation to calculate the vertical sag based on the catenary equation (only one fulcrum was fixed, whereas the other end was attached with a counterweight).

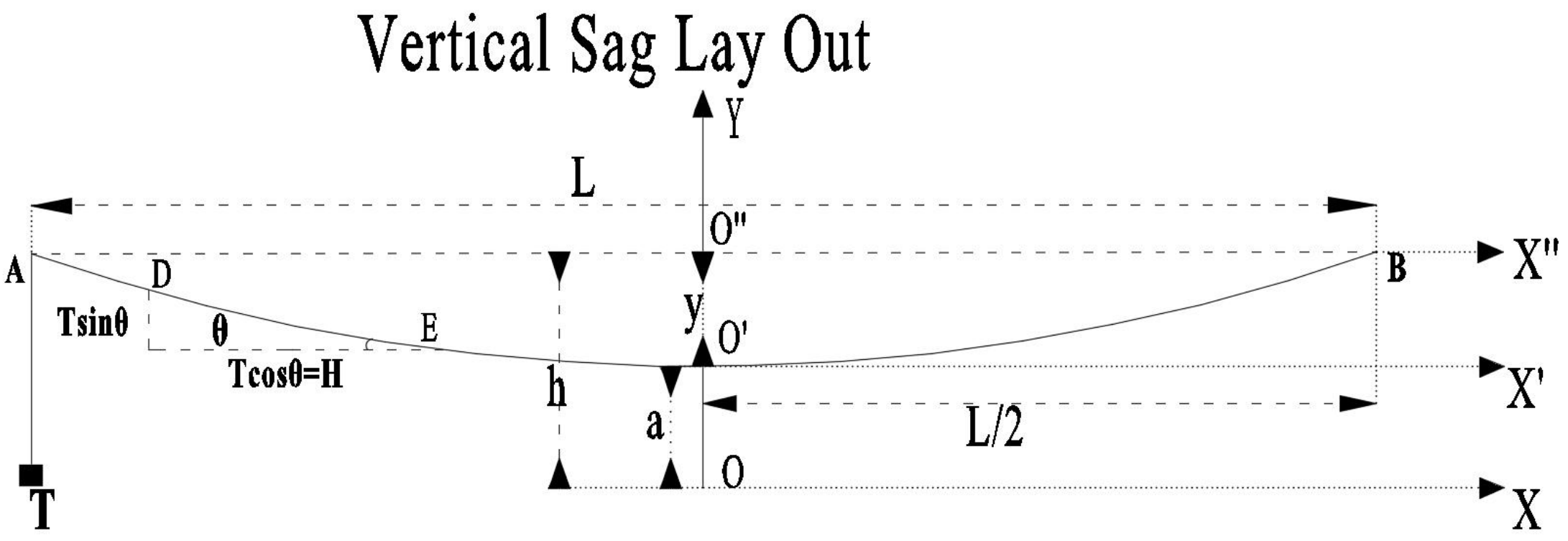

Figure 1 Layout of stretched wire

## 2.1 Catenary equation

Figure 1 shows that the left suspend point is A, the right suspend point is B, and the lowest point is O'. First, we set the hanging weight as $M$ (kg) and record the tension, where g is the acceleration due to gravity (N/kg), $L$ is the span length (m), and $y$ is the vertical sag. The force at point E was balanced. Next, we define a DE segment in the arc section of AO' and set the mass of the DE segment as $m$. Subsequently, E is subjected to an oblique upward pulling force $T$, and the angle between $T$ and the horizontal direction is $\theta$. The force analysis is as follows [21-23]: the horizontal component of the tensile force at point E is $H = Tcos\theta = Mgcos\theta = \mathrm{dx}$ , the vertical component of the tensile force at point E is $T\sin\theta = Mg\sin\theta = \mathrm{dy} = mg$ , and the differential equation can be expressed as

$$\frac{\mathrm{dy}}{\mathrm{dx}} = tan\,\theta = \frac{mg}{H} \tag{1}$$

In the equation above, $m = qs$ , where $q$ is the wire density (kg/m), i.e., the mass of the wire per unit length, and $s$ is the arc length of the wire DE segment. Subsequently, we substitute it into the differential equation as follows:

$$\frac{\mathrm{dy}}{\mathrm{dx}} = \frac{qsg}{H} \tag{2}$$

Next, we apply the Pythagorean Theorem as follows:

$$ds = \sqrt{dy^2 + dx^2} = \sqrt{1 + (\frac{dy}{dx})^2}dx \tag{3}$$

Consequently, we obtain the following equation:

$$s = \int \sqrt{1 + (\frac{dy}{dx})^2}dx = \int \sqrt{1 + y'^2}dx \tag{4}$$

Substituting Equation (4) into Equation (2) yields

$$\frac{dy}{dx} = \frac{qg}{H} \int \sqrt{1 + y'^2}dx \tag{5}$$

Next, variable substitution $p = \frac{dy}{dx} = y'$ is performed to obtain

$$p = \frac{qg}{H} \int \sqrt{1 + p^2}dx \tag{6}$$

To remove the integral sign, we take the derivative of $x$ on both sides of the equation above.

$$\frac{dp}{dx} = p' = \frac{qg}{H} \sqrt{1 + p^2} \tag{7}$$

Next, the variables are separated, and the two ends are integrated, as follows:

$$\int \frac{dp}{\sqrt{1 + p^2}} = \int \frac{qg}{H}dx \tag{8}$$

Because

$$\frac{d}{dx} sinh^{-1}(x) = \frac{1}{\sqrt{1 + x^2}} \tag{9}$$

The solution to the integral above is

$$sinh^{-1}(p) = \frac{qg}{H} x + C \tag{10}$$

In the equation above, $C$ is a constant whose value is determined as follows: When $x = 0, y'$

$= 0$, i.e., $p = 0$, this initial value condition is substituted into the solution obtained because $\sinh^{-1}(0) = 0$, and the solution is $C = 0$. Subsequently, the property of the inverse function is used to impose the hyperbolic sine on both sides of Equation (10) as follows:

$$p = y' = \frac{dy}{dx} = sinh(\frac{qg}{H} x) \tag{11}$$

The variables above were analyzed and integrated, as follows:

$$\int dy = \int sinh(\frac{qg}{H} x)dx \tag{12}$$

Consequently, the final solution was obtained, as follows:

$$y = \frac{H}{qg} cosh(\frac{qg}{H} x) + C \tag{13}$$

Because

$$a = H/_{qg} = Mgcos\ \theta/_{qg} = Mcos\ \theta/_{q} \tag{14}$$

The $C$ in the equation above is generally reserved, and it assumes different values depending on the coordinate system. When the arc is sufficiently small, i.e., $\theta$ is approximately 0, then $a = M/_{q}$. When $C = 0$, the standard equation of a catenary in a standard coordinate system (XOY) is obtained, as follows [25]:

$$y = a \cdot cosh(\frac{x}{a}) = \frac{M}{q} cosh(\frac{qx}{M}) \tag{15}$$

Where $x$ is the longitudinal length and its value is（$-L/2,L/2$）. Using the lowest point of the curve as the Y-axis zero point (the limit is $y(0) = 0$ ) when the horizontal axis is the X-axis, the coordinate system (X'O'Y) is established, and the equation of the contour catenary is as follows:

$$y(x, a) = a(cosh(\frac{x}{a}) - 1) = \frac{M}{q}(cosh(\frac{qx}{M}) - 1) \tag{16}$$

When we regard two fulcrum points of equal height as the Y-axis zero point (the limit is $y(-\frac{L}{2}) = y(\frac{L}{2}) = 0$ ) and the horizontal axis as the X-axis, we can establish a coordinate system (X''O''Y). Subsequently, the equation of the contour catenary is

$$y(x,a) = a(\cosh\frac{x}{a} - \cosh\frac{L}{2a}) = \frac{M}{q}(\cosh\frac{qx}{M} - \cosh\frac{Lq}{2M}) \qquad (17)$$

## 2.2 Vertical sag requirements

Because

$$\cosh(x/a) = \frac{1}{2}(e^{\frac{x}{a}} + e^{\frac{-x}{a}}) = \sum_{n=0}^{\infty}\frac{(\frac{X}{a})^{2n}}{(2n)!} = (1 + \frac{(\frac{X}{a})^2}{2!} + \frac{(\frac{X}{a})^4}{4!} + \cdots) \qquad (18)$$

Where $e = 2.71828$, when the quadratic term is disregarded, $x = L/2$ in the three coordinate systems above, and the sag of the lowest point of the catenary equations (Equations (15–17)) can be obtained as follows [10-21,26]:

$$|y| = q \cdot L^2/(8 \cdot M) = q \cdot g \cdot L^2/(8 \cdot T) \qquad (19)$$

In the equation above, $q = \rho\pi r^2$, where $r$ is the wire radius (m); $\rho$ is the wire unit volume mass（kg/m$^3$）, i.e., the bulk density (kg/m$^3$). It is noteworthy some studies [25-36] stated that the molecular term in Equation (19) should not be multiplied by the acceleration of gravity g, whereas other studies [37-39] considered it necessary. In fact, the multiplication of the acceleration by gravity depends on the units of $M$ in the denominator of Equation (19). If the units of $M$ is Newton instead of kilogram, then the numerator should be multiplied by the acceleration of gravity. The next subsections discuss the factors affecting the sag of a stretched wire.

### 2.2.1 Effects of hanging weight/tension and span length

Based on Equation (19), it is clear that the vertical sag is proportional to the wire density, diameter, and square of the span length, whereas it is inversely proportional to the applied pulling tension. Moreover, the vertical sag is positively correlated with the wire diameter and stretched length, whereas it is linearly positively correlated with the volume or linear density. As shown in Figure 2. A, under a fixed wire density, diameter, and length, the reduction in vertical sag was evident at the beginning (from 0.5 to 1 kg, the vertical sag decreased as the hanging weight/tension increased) but diminished later. Once the breaking force (1 kg) was reached, the wire broke (the break shown in the figure), rendering the subsequent reduction of the vertical sag impossible. As shown in Figure 2.b, under a fixed wire density, diameter, and hanging weight, the increase in the vertical sag was evident when the span length increased. In fact, if the vertical sag of the long-distance stretched wire exceeds 0.5 m (the break shown in the figure), then the layout and measurement accuracy of the wire will be severely affected [4].

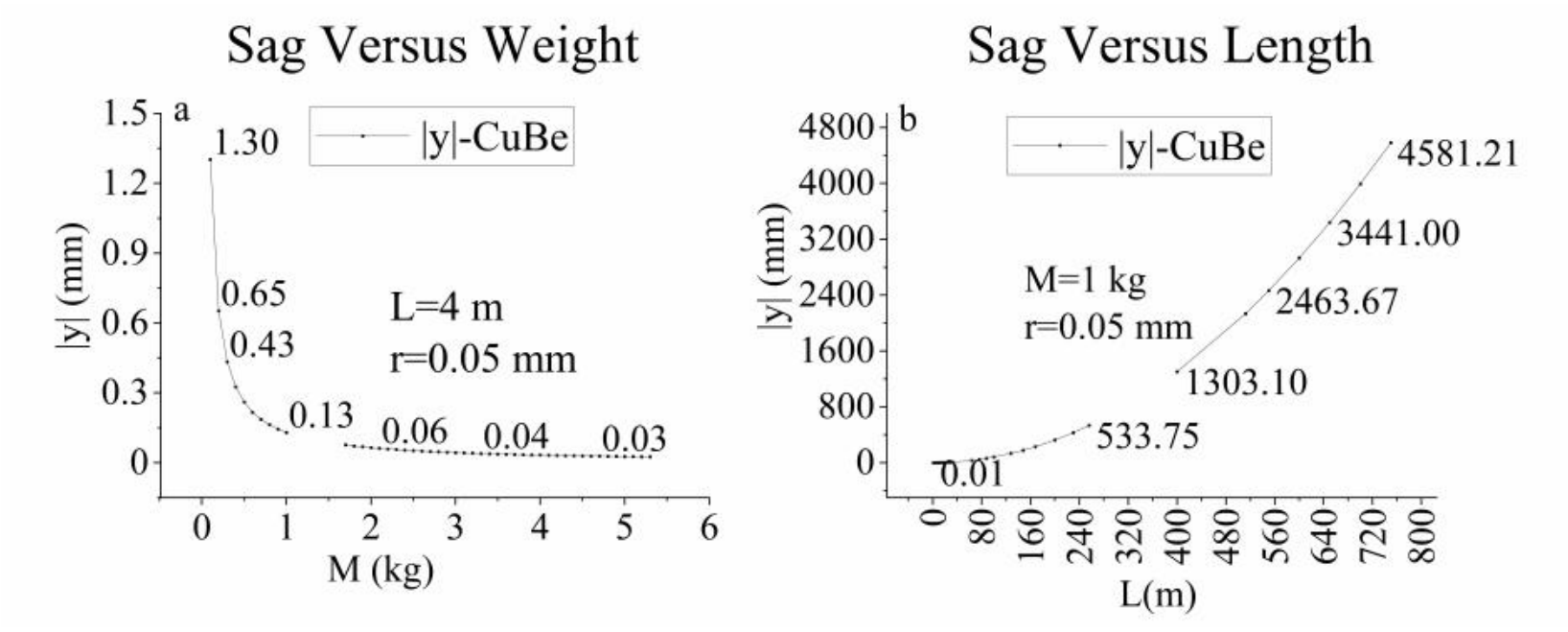

Figure 2 Relationships between vertical sag vs. counterweight and length

### 2.2.2 Effect of temperature change

Temperature change causes thermal expansion and contraction in the stretched wire body, manifested primarily as a change in the length of the wire body, thereby affecting the mass per unit length (q) [26-35]. Assuming that the external temperature is $t_0$, the expansion coefficient of the wire body is $\alpha$, and the total mass of the wire body is $m = q \cdot L$, when the external temperature increases and decreases $\Delta t$, the total length of the wire body changes $\Delta L$, and the wire is considered as an elastic body.

$$\Delta L = a \cdot L \cdot \Delta t \quad (20)$$

The effect of temperature change on the mass change per unit length of the wire is expressed as

$$\Delta q = {q \cdot a \cdot \Delta t} \Big/ {(1 + a \cdot \Delta t)} \quad (21)$$

Whereas the effect of temperature change on the vertical sag of the wire is expressed as

$$\Delta y_1 = {\pi \cdot r^2 \cdot \rho \cdot a \cdot \Delta t \cdot L^2} \Big/ {8 \cdot (1 + a \cdot \Delta t) \cdot M} \quad (22)$$

Finally, the relative sag deformation (in percentage) is expressed as

$$\Delta y_1 / y = a \cdot \Delta t / (1 + a \cdot \Delta t) \quad (23)$$

Because the largest linear expansion coefficients (typically used materials are stainless steel, indium steel, beryllium copper, nylon, carbon fiber, and other synthetic fibers, as detailed in Table 1) are on the order of $10^{-4}$ (units: m), and if the maximum temperature change is -300 °C (in the low-temperature superconducting cryostat), then the relative sag deformation is -2.9%. If the maximum expansion coefficient is less than $10^{-5}$ (units: m), then the relative sag is less than -0.299%.

### 2.2.3 Effect of elastic deformation

The elastic deformation of the tension wire affects its position measurement. The main effect of tension is to reduce the mass per unit length, i.e., the linear density [1,40,41]. The decrease in wire density causes a decrease in the wire sag.

$$\Delta y_2 = 1/(1+\varepsilon) \cdot q \cdot L^2 / 8M = 1 / (1 + T/s \cdot E) \cdot q \cdot L^2 / 8M = \pi \cdot \rho \cdot r^2 \cdot L^2 / 8 \cdot M \cdot (1 + Mg / s \cdot E) \quad (24)$$

Where $\varepsilon$ is the strain (%), $E$ the modulus of elasticity (MPa = N/mm$^2$), and $s$ the cross-sectional area (mm$^2$). The percentage of wire deformation on the vertical sag is expressed as

$$\Delta y_2 / y = 1 / (1 + Mg / s \cdot E) \quad (25)$$

Except for nylon, the elastic moduli of the other materials listed in Table 1 are within 100–200 GPa. If the diameter of the wire is 0.1–0.2 mm and the wire tension is regarded as the breaking strength, then the percentage of wire elastic deformation on the vertical sag is 0.2%–0.3%

### 2.2.4 Effect of Earth's rotation

It is well known that the rotation of the Earth will cause gravitation [41] (and inertial force [42], also known as the deflection force of the Earth's rotation or Coriolis force). The authors of [42] qualitatively analyzed the effect of universal gravitation on mechanical tension wires and considered East–West as the best direction for tension wires, and the North–South as the most unfavorable. When the observation frequency is half a month, it is best to select 0 o'clock and 12 o'clock on the first and fifteenth of each month, respectively. If the observation frequency is one month, then it is best to select 0 o'clock and 12 o'clock on the fifteenth of each month. In [42], the effect of inertial force on a mechanically stretched wire was quantitatively analyzed, and it was believed that inertial force would reduce the change in its vertical deviation and result in a horizontal deviation (even without side wind effects). The higher the latitude, the stronger was the inertial force. In theory, the inertial force does not exist on the equator. The inertial force in the vertical direction will impose the following effects on the vertical sag:

$$\Delta y_3 = \pi \cdot \rho \cdot r^2 \cdot L^2 \cdot \omega^2 \cdot R \cdot cos_\phi^2 \Big/ 8 \cdot M \tag{26}$$

Where $\omega$ is the Earth's rotation angular velocity ($7.25 \times 10^{-5} rad/s$), $\varphi$ the latitude of the tension line (used the angle of the integer, Lanzhou, Gansu, China N 36°), and $R$ is the Earth's radius (6371 km, used in meters). As shown from the equation, the effect of inertial force on the vertical sag is proportional to the angular velocity of the Earth's rotation, the radius of the earth, the density of the wire, and the length of the wire, whereas it is inversely proportional to the dimension and tensile force. Because the angular velocity of the Earth's rotation is relatively stable, the difference is from the latitude of the tension line. To evaluate the relative effect of the inertial force on the vertical sag change, the percentage can be used to evaluate

$$\Delta y_3 \big/ y = \omega^2 \cdot R \cdot cos_\phi^2 \big/ g \approx 3.417 \times 10^{-3} cos_\phi^2 \tag{27}$$

A study by Conseil Européenn pour la Recherche Nucléaire （CERN） showed [42] that the percentage of the effect of inertial force on vertical sag change was 0.2%. When measuring the accelerator alignment, if the vertical deviation can be controlled to within 40 µm, then the change in the vertical sag will be within 0.08 um, which is negligible.

# 3 Experimental section

### 3.1 Wire material and its properties

Although stainless steel wires of diameter 0.5–1.2 mm are typically used in water conservancy and hydropower deformation monitoring, they are not suitable for the high-precision baseline measurements of accelerator engineering owing to their significant vibration amplitudes, sagging, and diameter changes. Beryllium copper wire of diameter 0.1–0.5 mm is widely used in accelerator alignment measurement; although it exhibits high-frequency resonance, its amplitude is extremely small [15]. In addition, Be–Cu wire exhibits good conductivity and a low stress density ratio ($\leqslant 0.2 \times 10^6$ N.m/kg) and can be used as a short-distance stretched wire. Carbon fiber and its reinforced composite materials (e.g., carbon fiber reinforced polymer, CFRP) and synthetic materials (e.g., dipropyl phthalate reinforced polymer, DPRP) have high stress–density ratios (⩾2 × $10^6$ N·m/kg) and can be used to lay tension wires over long distances; however, wires manufactured using such materials have high electrical impedance and low conductivity, and hence not suitable for direct conductors. Among them, nylon filaments and Kevlar are nonconductive.

Tab. 1 Properties of different materials

| Wire Material | Density (kg/m$^3$) | Elastic Modulus (GPa) | Tensile Strength (MPa) | Linear expansion coefficient (m/m.°C) | Strength-to-density ratio ($10^6$ N.m/kg) | Conductivity ($10^6$ s/m) at 20 °C |
|---|---|---|---|---|---|---|
| 304 | 7930 | 193 | 520 | $1.7\times10^{-5}$ | 0.07 | 1.37 |
| Be–Cu | 8300 | 128 | 1300 | $1.8\times10^{-5}$ | 0.16 | 13.1 |
| Indium | 8100 | 160 | 590 | $1.6\times10^{-6}$ | 0.07 | 11.9 |
| Nylon | 1200 | 1.4–5.2 | 47–110 | $1–5\times10^{-4}$ | 0.04 | - |
| CFRP | 1500–1800 | 125–-181 | 2500–4200 | $7\times10^{-7}$ | 2.11 | 0.02 |
| KEVLAR | 1440 | 71–112 | 2757 | $2.5\times10^{-5}$ | 1.91 | - |
| DPRP | 1500 | 140 | 3200 | $1\times10^{-6}$ | 2.13 | - |

## 3.2 Experimental method and measurement process

As shown in Figure 3, in the experiment, one end was fixed, and the other end was suspended with a counterweight to achieve tension. The length of the stretched wire was 4 m, and the heights of the ruler and stretched wire were adjusted to the same height using a level. The Keuffel and Esser paragon tilting level can realize both longitudinal distance and vertical sag measurements with the aid of a ruler. When we used the level for observation, the readings were recorded once every 0.1 m in the longitudinal direction. The median of three observations instead of the average was used to improve the accuracy and robustness. The wire body materials used in the experiment were 304 stainless steel and Be–Cu of radii 0.05 and 0.1 mm, respectively.

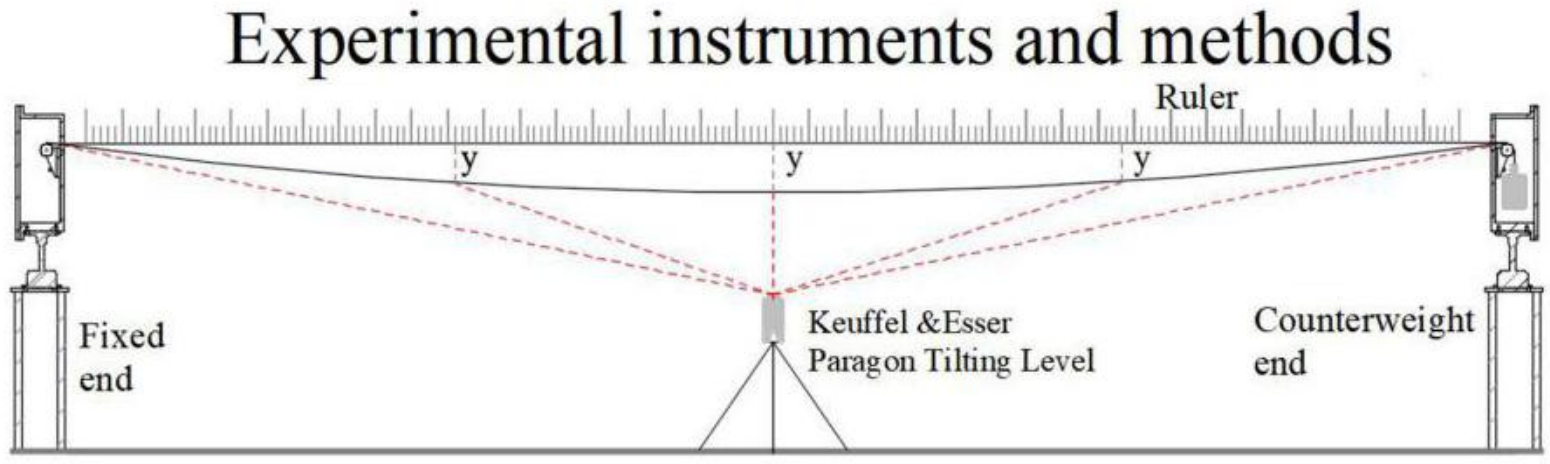

Figure 3 Illustration of experimental scheme

Based on the author’s experience, when the pulling force of 304 stainless steel and Be–Cu wire is close to the break force, the beginning of the stretch is the most difficult. Hence, caution is required to ensure that the wire will not break when subjected to the break force. However, once the wire is stretched successfully, the material can be used stably for some time even if it bore the breaking tension. When stretching materials in such a manner, it is necessary to ensure that the wire is not bent to achieve a smooth stretch. Furthermore, it is preferable to pre-stretch 50% of the break force and then apply 100% of the break force.

The Keuffel & Esser paragon tilting level comprises a telescope with an effective aperture of 30 mm, an optical micrometer unit with $\pm 100$ intervals, a coincidence reading bubble, a circular bubble, a focusing knob with magnification from 20 to 30, and leveling screws. The level has a minimum focus of 101.6 mm. Its optical micrometer has a survey range of $\pm 2.54$ mm and a direct reading accuracy of 0.0254 mm, which can be estimated to be 0.00254 mm. Even if the level is set appropriately, errors can never be completely eliminated from our setup. To ensure measurement accuracy, the level should be regularly sent to the certified national metrology department for calibration before its use and calibrated as follows. First, the circle bubble should be verified and calibrated to guarantee that its axis is parallel to the vertical axis of the instrument. Subsequently, the coincidence split-bubble should also be verified and calibrated to guarantee that its axis is parallel to the collimated sight axis; next, the horizontal wire of the reticule should be verified and calibrated to guarantee that the horizontal wire is perpendicular to the instrument’s vertical axis; finally, the atmospheric refraction and Earth curvature should be calibrated. When using the level, the circular bubble should be adjusted to its dead center using three thumb screws, and both the left and right splits should be aligned by adjusting the tilting screw.

# 4 Results

Considering temperature change, the elastic deformation of the wire body and the Earth's rotation, the vertical sag is calculated as follows:

$$y_{\text{unltimate}} = y + \Delta y_1 + \Delta y_2 + \Delta y_3 = -\left(1 - \frac{\alpha \cdot \Delta T}{1 + \alpha \cdot \Delta T} - \frac{1}{1 + M / (s \cdot E)} - \omega^2 \cdot R \cdot \cos_{\phi}^2\right) q \tag{28}$$

As shown in Figure 4.a, the horizontal axis of the sag of 304 stainless steel is 29 times smaller and its vertical axis is 86 times larger than the actual size. As shown in Figure 4.b, the horizontal axis of the cube's sag is 29 times smaller, and its vertical axis is 208 times larger than the actual size. The upper number of each picture represents the distance from the stretched wire to its center, the middle number represents the theoretical sag of each point, and the lower part represents the three sag observations of each point. In Figure 4, the observation value multiplied by 0.0254 is the elevation expressed in millimeters; the two sides of the bottom are the datum elevation; the middle observation value of the bottom minus the datum height of both sides multiplied by 0.0254 is the elevation difference in millimeters.

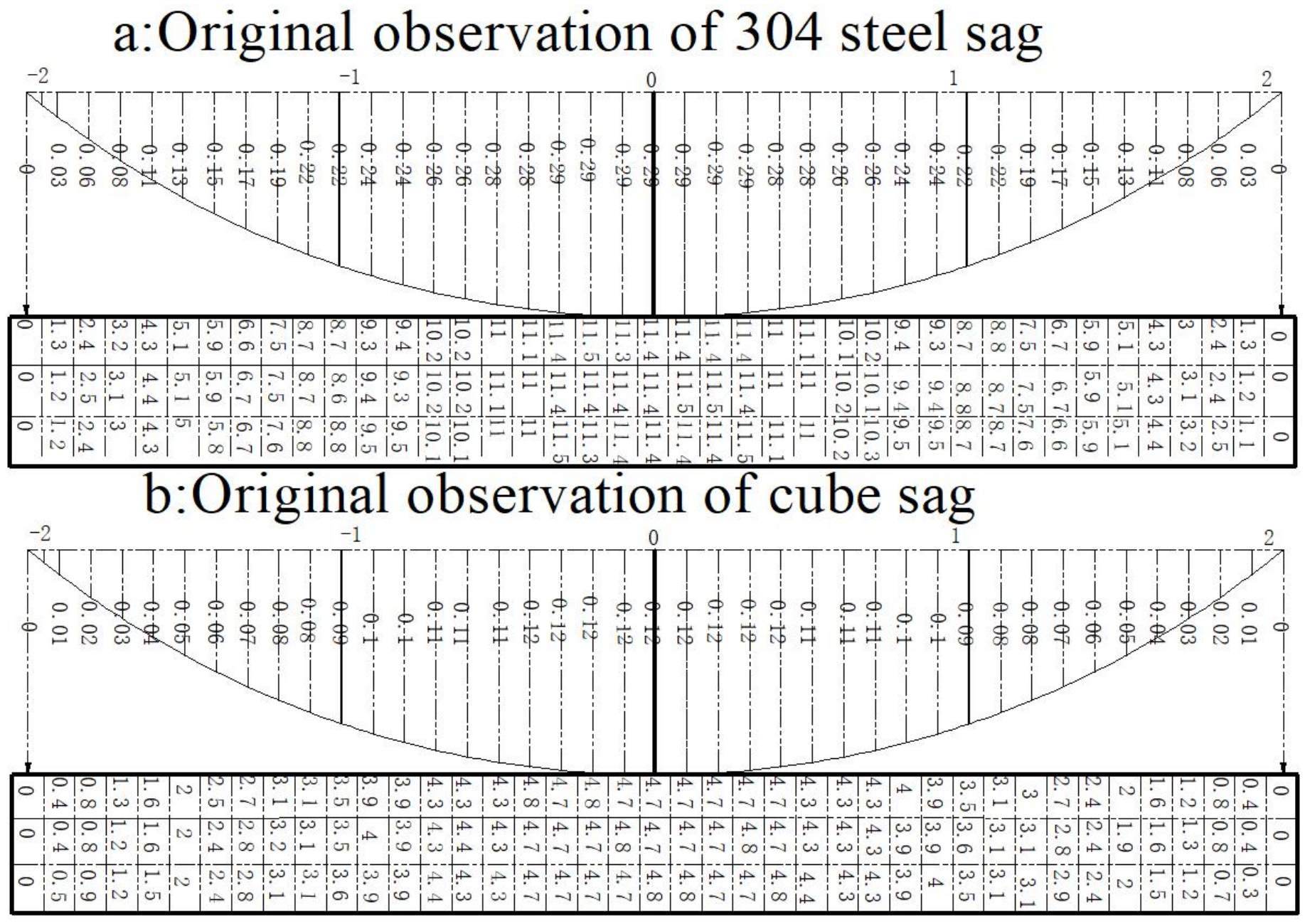

Figure 4 Original observation of sag

Figures 5.a and b show the theoretically calculated and actual measured values of vertical sag of the 304 stainless steel at different radii (0.05 and 0.1 mm, respectively). Figures 5.c and d show the theoretically calculated value and actual measured values of vertical sag of the Be–Cu wire at different radii (0.05 and 0.1 mm, respectively). The stretched wire length was 4 m, and the maximum tensile force was applied. As shown by the actual results in the figure, the theoretical calculation of the vertical sag was consistent with the actual measurement. Furthermore, the result shows that under a certain length, the vertical sag is the same for a wire of a different radius at the maximum tension.

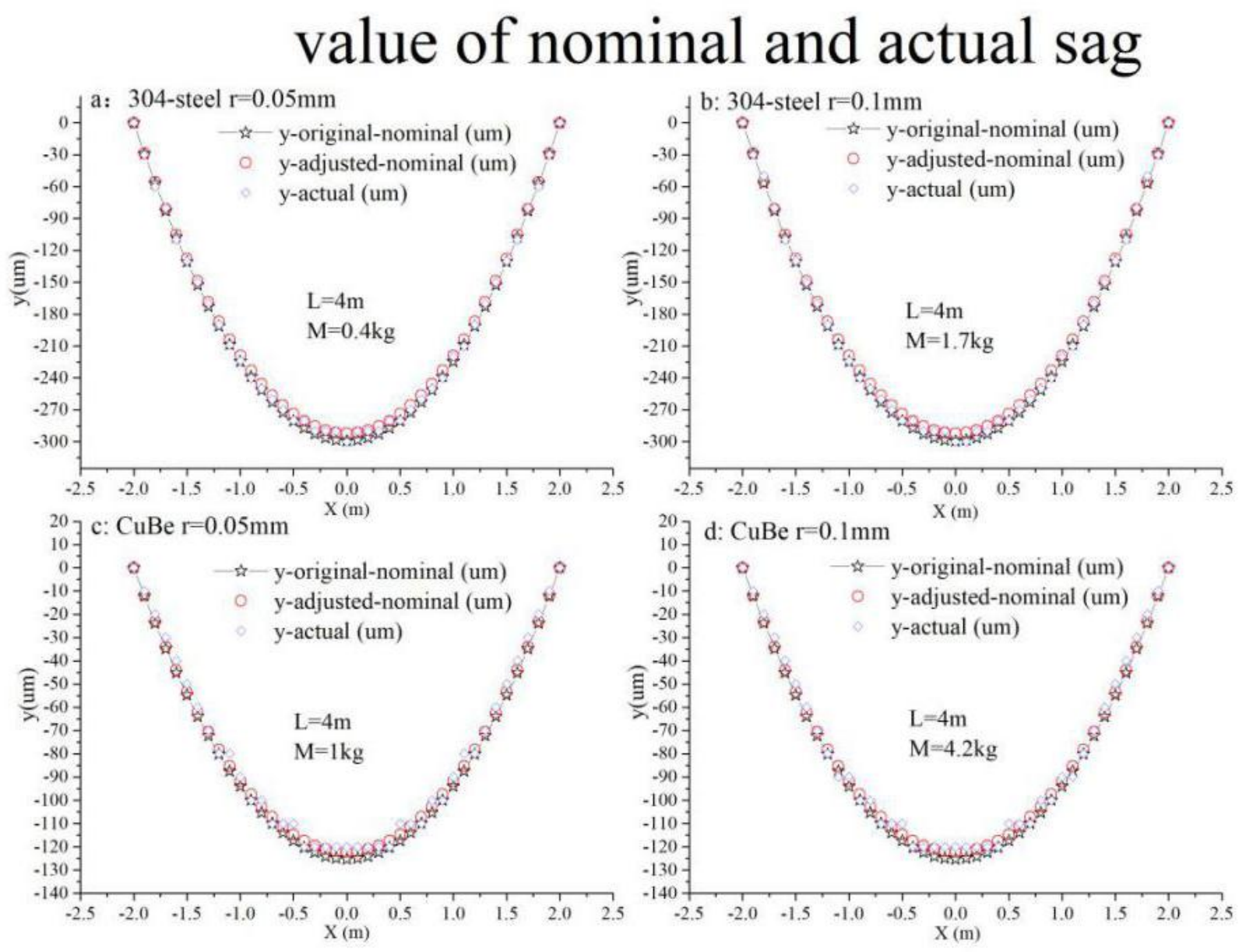

Figure 5 Calculated and measured values of vertical sag

The original and adjusted nominal sag shown in Figure 5 were computed using Equations (19) and (28), respectively. If the maximum temperature change is ±300 °C, then the maximum sum of the effects of temperature change, elastic deformation, and Earth rotation is between -2.4% and -3.4% of the nominal sag. Therefore, their effects on the vertical sag can be disregarded based on the negligence principle.

# 5 Discussion

As shown in Figure 2 and Section 2.2.1, a small sag can be obtained by increasing the hanging tension/weight at the beginning; however, this is not applicable near the break sag (the break shown in Figure 2). Therefore, the effects of the hanging tension/weight, specific strength, and length are discussed below.

## 5.1 Effect of maximum hanging tension/weight

The maximum hanging tension $M_{max}$ is the tensile force $F_{max}$ ($F_{max} = M_{max}g$). To minimize the vertical sag, the maximum tensile force should be smaller than the break strength (tensile strength) of the brittle material, i.e., the hanging break weight should be

$$M_{\max} \leq s \cdot [\sigma] / g = \pi \cdot r^2 \cdot [\sigma] / g = \pi \cdot r^2 \cdot R_m / (n \cdot g) \tag{29}$$

,

Where $s$ is the wire cross-sectional area (mm$^2$), $R_m$ refers to the tensile strength is the maximum load that a material can support without fracture when being stretched (MPa) and $[\sigma]$ is the allowable stress (MPa). The relationship between the allowable stress and the break or tensile strength is expressed as

$$[\sigma] = R_m / n \tag{30}$$

Where $n$ is the safety factor between 1 and 2. When it is 1, the maximum hanging weight is

$$M_{max} \leq \pi \cdot r^2 \cdot R_m / g \tag{31}$$

To avoid the wire from breaking, the maximum tensile force should be less than 50%–67% of the allowable stress. For example, when $n$ is 1.5, the hanging weight is

$$M_{\text{actual}} = 0.67 \cdot \pi \cdot r^2 \cdot R_m / g \tag{32}$$

The equation above shows that the maximum allowable tensile force is proportional to the tensile strength and the square of the material radius or diameter. Once the material is selected, the tensile strength of the material becomes constant, and it can be concluded that to increase the allowable tensile force, its diameter or radius must be expanded.

**5.2 Effect of specific strength and length**

After the vertical sag of the stretched wire is determined, one must ensure that the pulling force is less than the maximum tensile force (substituting Equation (29) into Equation (19)); the vertical sag requirement is

$$y \geq \frac{\pi \rho r^2 L^2 ng}{8 \pi r^2 R_m} = \frac{ng \rho L^2}{8 R_m} \approx 1.226 \frac{nL^2}{R_m / \rho} \tag{33}$$

When the safety factor $n$ is 1, then the break vertical sag is

$$y_{\text{break}} = \frac{\pi \rho L^2 r^2 g}{8 \pi r^2 R_m} = \frac{g \rho L^2}{8 R_m} \approx 1.226 \frac{\rho L^2}{R_m} \approx 1.226 \frac{L^2}{R_m / \rho} \tag{34}$$

Based on the Equations (33) and (34), which are derived from Equation (19) of Section 2, the vertical sag of the stretched wire is determined by the specific strength of the material (i.e., the strength-to-density ratio) and the square of its length. The similarity between the two equations is that they both consider the length and density. The difference between Equations (34) and (19) is that Equation (34) indicates the sag's essential and real factors (such as the safety factor and tensile strength of the wire material), whereas Equation (19) only indicates the surface factors (such as the hanging tension/weight). Because the safe tension/weight of the wire material was exceeded, wire breakage occurred, as shown in Figure 2.

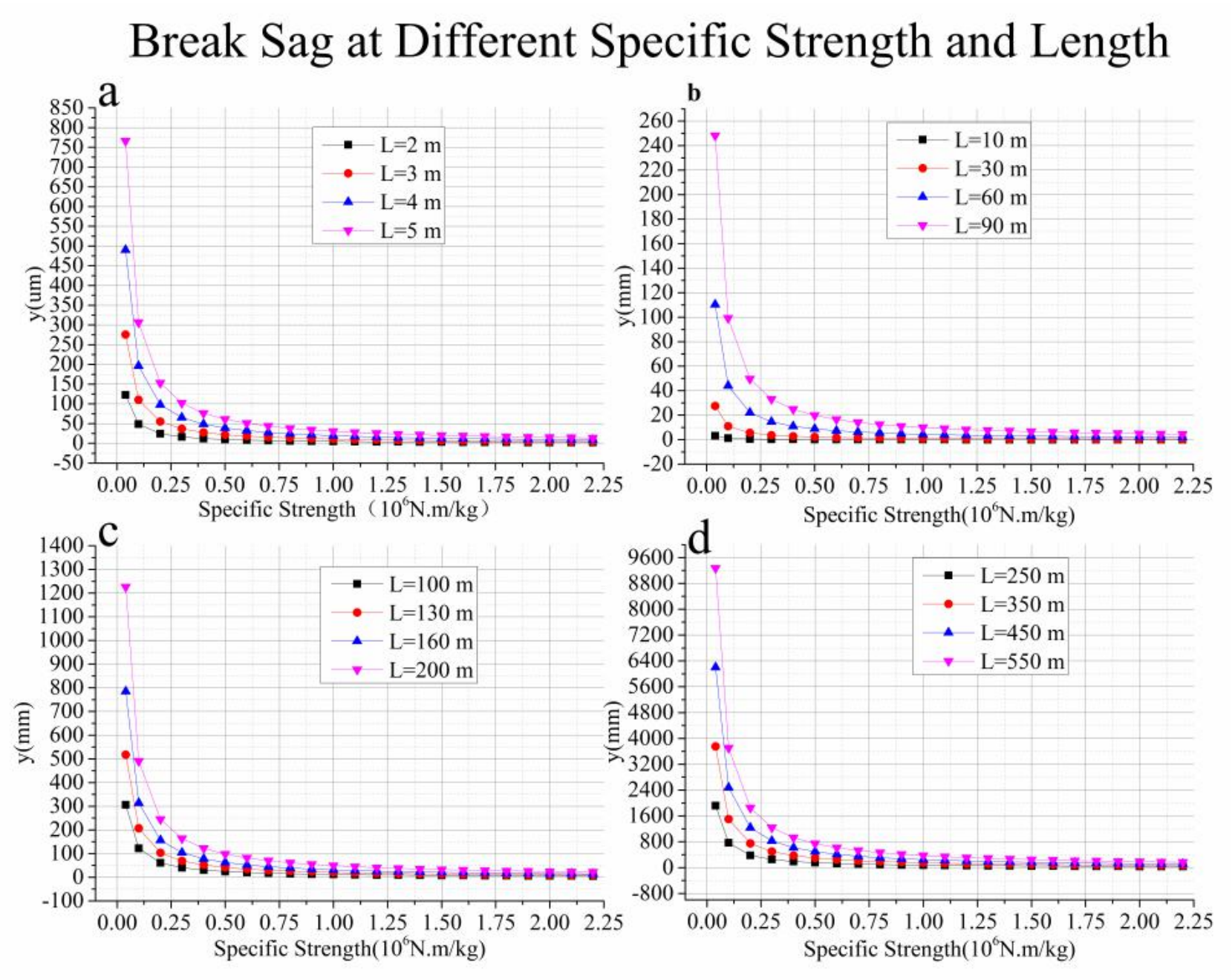

Figure 6 Relationship between vertical sag and specific strength and length

As shown in Figure 6, if the vertical sag is required to be small (e.g., the vertical sag was required to be less than or equal to 20 μm in [1]), a material with sufficient specific strength (such as carbon fiber and other materials) can be selected. However, the conductivity of carbon fiber without post-treatment is relatively poor, and it is not suitable for both conductors and stretched wires. Although carbon fiber has been treated with technologies such as silver plating on its surface, its conductivity can be improved significantly. However, only ordinary carbon fibers are easily available commercially. The ideal vertical sag should be 1.5–2 times the fracture vertical sag to ensure that the wire is stretched perfectly without causing it to break, as well as to enable long-term deformation monitoring. It has been reported [1] that in the ultra-high-precision alignment of accelerators (e.g., the calibration of beam position monitors, various types of magnets, and magnetic field measurements within a range of less than 2 m), the vertical sag of the stretched wire should be less than 20 μm, but the vertical sag requirements must be consistent with the properties of the actual material not only to ensure the minimum sag, but also to ensure that the wire does not break. Subsequently, according to the pulling force, $M_{actual} \leq M_{max}$, and the vertical sag suitable for the material is determined by combining the length of the stretched wire and the specific strength of the material.

## 6 Conclusion

The vertical sag of a stretched catenary depends on the specific strength of the wire material and its stretched length; additionally, it is not directly related to the diameter (radius) of the wire. The safety factor n is 1.5 or beyond, and the stretched wire can operate stably. The allowable tensile force is determined by the breaking strength or tensile strength of the wire material and the wire diameter or radius.

## Acknowledgement

The authors thank the two anonymous reviewers who have helped to improve the manuscript significantly. The work was supported by Large Research Infrastructures "China initiative Accelerator Driven System"(Grant No.2017-000052-75-01-000590 ).

All authors contributed to the study conception and design. Material preparation, data collection and analysis were performed by Jiandong Yuan and Xudong Zhang. The first draft of the manuscript was written by Jiandong Yuan and all authors commented on previous versions of the manuscript. All authors read and approved the final manuscript.